\DocumentMetadata{}
\documentclass[sigconf]{acmart}
\usepackage{hyperref}
\usepackage{hyperxmp}

\AtBeginDocument{%
  }

\setcopyright{acmcopyright}
\copyrightyear{2025}
\acmYear{2025}
\acmDOI{XXXXXXX.XXXXXXX}

\acmConference[EASE 2025]{The 29th International Conference on Evaluation and Assessment in Software Engineering}{17–20 June, 2025}{Istanbul, Türkiye}
\acmPrice{00.00}

\usepackage{amsmath}
\usepackage{graphicx}
\usepackage{textcomp}
\usepackage{colortbl}
\usepackage{enumitem}
\usepackage{array,booktabs}
\usepackage{multirow}
\usepackage{balance}
\usepackage[framemethod=default]{mdframed}
\usepackage[normalem]{ulem}

\usepackage{array}
\usepackage{makecell}
\usepackage{tcolorbox}

\newcommand{\PreserveBackslash}[1]{\let\temp=\\#1\let\\=\temp}
\newcolumntype{C}[1]{>{\PreserveBackslash\centering}p{#1}}
\newcolumntype{R}[1]{>{\PreserveBackslash\raggedleft}p{#1}}
\newcolumntype{L}[1]{>{\PreserveBackslash\raggedright}p{#1}} 

\begin{document}

\title[Assessing the Bug-Proneness of Refactored Code: Longitudinal Multi-Project Studies]{Assessing the Bug-Proneness of Refactored Code:\\A Longitudinal Multi-Project Study}

\author{Isabella Ferreira}
\orcid{0000-0002-9884-5890}
\affiliation{%
  \institution{Pontifical Catholic University of Rio de Janeiro}
  \city{Rio de Janeiro}
  \country{Brazil}
}

\author{Lawrence Arkoh}
\orcid{0009-0005-5904-9313}
\affiliation{%
  \institution{North Carolina State University}
  \city{Raleigh}
  \country{U.S.A}
}

\author{Anderson Uchôa}
\orcid{0000-0002-6847-5569}
\affiliation{%
  \institution{Federal University of Ceará}
  \city{Itapajé}
  \country{Brazil}
}

\author{Ana Carla Bibiano}
\affiliation{%
  \institution{Pontifical Catholic University of Rio de Janeiro}
  \city{Rio de Janeiro}
  \country{Brazil}
}

\author{Alessandro Garcia}
\orcid{0000-0001-5788-5215}
\affiliation{%
  \institution{Pontifical Catholic University of Rio de Janeiro}
  \city{Rio de Janeiro}
  \country{Brazil}
}

\author{Wesley K. G. Assunção}
\orcid{0000-0002-7557-9091}
\affiliation{%
  \institution{North Carolina State University}
  \city{Raleigh}
  \country{U.S.A}
}

%

\definecolor{cellh}{gray}{0.7} 

\renewcommand{\shortauthors}{Surename et al.}

\newcommand{\la}[1]{{\textcolor{magenta}{~[~\textbf{LA}: \textit{#1} ]}}}
\begin{abstract}
Refactoring is a common practice in software development, aimed at improving the internal code structure in order to make it easier to understand and modify. Consequently, it is often assumed that refactoring makes the code less prone to bugs. However, in practice, refactoring is a complex task and applied in different ways (e.g., various refactoring types, single vs. composite refactorings) and with a variety of purposes (e.g., root-canal vs. floss refactoring). Therefore, certain refactorings can inadvertently make the code more prone to bugs.
Unfortunately, there is limited research in the literature on the long-term relationship between the different characteristics of refactorings and bugs. 
This paper presents a longitudinal study of 12 open source software projects, where 27,450 refactorings, 6,051 reported bugs, and 49,250 bugs detected with static analysis tools were analyzed.  
While our study confirms the common intuition that refactored code is less bug-prone than non-refactored code, we also extend or contradict existing body of knowledge in other ways. First, a code element that undergoes multiple refactorings is not less bug-prone than an element that undergoes a single refactoring. A single refactoring is the one not performed in conjunction with other refactorings in the same commit. Second, single refactorings often induce the occurrence of bugs across all analyzed projects. Third, code elements affected by refactorings made in conjunction with other non-refactoring changes in the same commit (i.e., floss refactorings) are often bug-prone. Finally, many of such bugs induced by refactoring  cannot be revealed with state-of-the-art techniques for detecting behavior-preserving refactorings.

\end{abstract}



\maketitle

\section{Introduction}
\label{sec:introduction}

Software maintenance requires the application of various changes to the source code~\cite{kemerer1995, benestad2009}. Undisciplined changes often lead to the degradation of the code structures~\cite{46, 50}. Such degradation potentially hinders the software's maintainability~\cite{coutinho2022influential}. One way to observe the code structure degradation is through \textit{code smells}, which are poor design or implementation choices in the program that represent symptoms of a structural problem~\cite{19,oliveira2022developers}. To address these symptoms, developers often apply code \textit{refactorings}~\cite{6, 19}. Code refactoring is a transformation aiming at improving the program structure while preserving its observable behavior~\cite{19}. 
Refactoring is a complex task for several reasons as we discuss in the following. 

First, developers apply refactoring with different purposes~\cite{33, 34}, such as making the code easier to understand/modify and improve program testability (a.k.a., root-canal refactoring); or facilitating feature additions and even supporting bug fixes (a.k.a., floss refactoring)~\cite{33, 34, 35}. Developers apply \textit{root-canal (or pure) refactoring} when they aim to exclusively improve the code structure quality. Otherwise, developers apply \textit{floss refactoring} when they aim to refactor the code together with non-structural changes as a means to reach other goals~\cite{18, 56}.
Second, applying certain \textit{refactoring types} is very challenging~\cite{40, 57}: for example, due to inter-component dependencies there is a need for changing different places of the large code bases (e.g., parameters, methods, or classes)~\cite{40, Justin2024}. 
Third, sometimes developers apply \textit{single refactorings} to improve a code structure, but there are cases that require \textit{composite refactorings} (i.e., a sequence of multiple refactorings) to remove a code smell~\cite{bibiano2023composite}. However, composite refactorings can eliminate one code smell, but if not applied carefully, create another smell even more harmful~\cite{bibiano2021look}.
Fourth, developers acknowledge that it is difficult to ensure program correctness after refactoring~\cite{40}. In a study conducted at Microsoft~\cite{40}, 76\% of developers mentioned that refactoring comes with the risk of introducing bugs and functionality regression. 
Finally, to make matters worse, although there exist refactoring tools, developers often refactor code manually, because of limitations and lack of flexibility of such tools~\cite{eilertsen2021usability,oliveira2023untold}.

Despite the goal of making the code easier to understand and modify, given the complexity of refactoring in practice, its application might negatively impact the code. Such impact can increase the \textit{bug-proneness} of refactored code elements, making the code structure more susceptible to containing a bug in the future.  
The susceptibility of a refactored code element may depend on particular refactoring characteristics. In this paper, we particularly conjecture that the refactoring type, the purpose of the refactoring (root-canal vs. floss), the number of refactoring (single vs. composites) can influence on the bug proneness of the refactored code element. In summary, we address a prevailing question in the practice of software refactoring: \textit{What refactoring characteristics make the refactored code more susceptible to containing bugs?}

Unfortunately, the bug-proneness of refactored code has received little attention in the literature. Only a few studies analyze the relation between refactored code and bugs~\cite{1, 69, 67, 68, 3}. Interestingly, these studies indicate that refactored code is often susceptible to containing bugs, which contradicts the main goal of refactoring. However, these studies are limited in several ways. First, they do not provide clear, in-depth evidence linking bugs to refactoring. Second, current studies lack an understanding of how refactoring impacts the bug-proneness of refactored elements, and how long it takes for the refactored code element to become buggy. Third, there is limited evidence of whether refactoring characteristics (e.g., various refactoring types, root-canal vs. floss, and single vs. composites) reduce or increase the bug-proneness of refactored code.

The goal of this work is to perform a fine-grained and systematic analysis of refactored code to better understand its bug-proneness over time. 
Thus, we present a longitudinal study of 12 open-source projects. Differently from existing studies, we analyze a long period of historical data for these projects, which consist of 27,450 refactoring operations, 6,051 reported bugs, and 49,250 bugs detected with static analysis tools. We analyze to what extent refactored and non-refactored code elements are susceptible to have bugs, the impact of refactoring types, the difference in bug proneness for single and composite refactorings, and the impact of different refactoring purposes on making a code element become buggy. Also, our analysis verifies the distance of refactoring operations and the occurrence of bugs in later versions. The distance is computed by the number of changes that separates the refactored version of a program element and the buggy version (if any) of the same element. 

Our study confirms the common intuition that refactored code is less prone to bugs than non-refactored code. We provide recommendations to practitioners and researchers based on our findings, and extend or contradict the existing  knowledge in other ways:

\begin{itemize}
    \item We contradict the findings of the aforementioned studies (e.g.,~\cite{3}). Our study shows that a code element that undergoes composite refactorings is as bug prone as an element that undergoes a single refactoring. We found that the single refactoring tends to be closer to becoming buggy than the composite refactorings.
    
    \item Many single operations of root-canal (i.e. pure) refactoring induce the occurrence of bugs across all projects. Surprisingly, these operations were mostly instances of simple program transformations, such as Extract Method or Inline Method. We did not observe the bug-inducing refactoring patterns documented by Bavota et al.~\cite{1}. Our data set is larger than theirs, and we analyzed all the program versions instead of only its major  releases.
    
    \item Code elements affected by root-canal refactorings are often bug prone. Therefore, independently of the complexity of certain refactoring edits, developers unconsciously induce bugs through refactoring.
    
\end{itemize}

\section{Motivating Example and Problem Statement}

This section presents a real case of refactored code that originates a bug. The refactoring of our example took place to add a new feature (floss refactoring). This is a kind of example that evidences the need of investigating different characteristics of refactoring on bug proneness. Then, we highlight the practical cases that are not investigated in existing literature, motivating our work.

\subsection{Bug-Proneness of Refactored Code}
\label{sec:motivating-example}

\begin{figure}[!tp]
    \centering
    \includegraphics[width=1\linewidth]{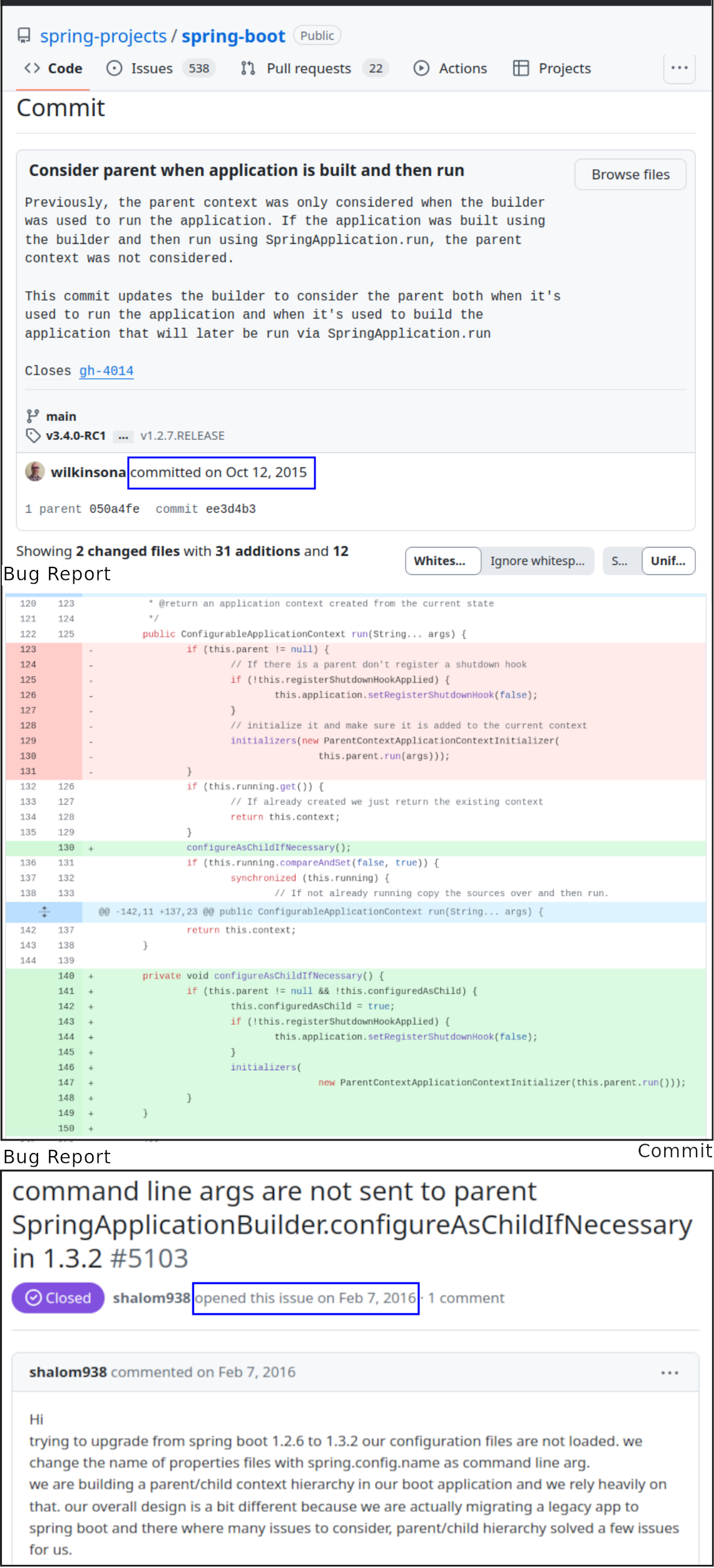}
    \caption{Example of a bug originated from an Extract Method}
    \label{fig:motivating-example}
\end{figure}

\sloppy
We motivate our study by illustrating to what extent refactored code could be bug-prone~\cite{1, 3}. This example is taken from the Spring Boot\footnote{\url{https://github.com/spring-projects/spring-boot.git}} project on GitHub, and presented in Figure~\ref{fig:motivating-example}. 
We analyzed the bug report \#5103\footnote{\url{https://github.com/spring-projects/spring-boot/issues/5103}}, which was open on Feb 7, 2016, as shown at the bottom of Figure~\ref{fig:motivating-example}. The description of the bug report mentions that the configuration file is not loaded after upgrading Spring Boot from version 1.2.6 to 1.3.2. The reporter of this bug finds that in version 1.3.2, a method named \texttt{configureAsChildIfNecessary} calls \texttt{ParentContextApplicationContextInitializer} without passing the command line arguments. The bug was reported not long after the release Spring-boot 1.3.2. 

We then analyzed the code associated with the origin of this bug and noticed that an Extract Method refactoring was applied to the code element four months before the bug was reported. As presented at the top of Figure~\ref{fig:motivating-example}, in the Commit ee3d4b3\footnote{\url{https://github.com/spring-projects/spring-boot/commit/ee3d4b3}} on Oct 12, 2015; the lines of code 123-131 were extracted from the method \texttt{run} to a new method called \texttt{configureAsChildIfNecessary}. However, we can observe in the code extracted, that the arguments are not passed to the new method. Going further in the analysis, we observed that the Extract Method refactoring was meant to introduce a new feature as part of a new version of Spring-boot. Thus, this bug originated with floss refactoring.

This illustrative case highlights bug proneness of refactorings, which can be affected in cases such as floss refactorings. This motivates our work, on exploring the bug proneness of refactored code in the context of various characteristics of refactorings.

\subsection{Problem Statement}


Developers might unconsciously make the source code more susceptible to containing bugs, depending on the complexity of certain refactorings (see the discussion of refactoring complexity in Section~\ref{sec:introduction}). Kim et al.~\cite{40} found that there is no safe way of checking refactoring correctness, mainly when a regression test suite is insufficient. As a result, regression bugs might be introduced to the source code. Consequently, developers prefer to avoid refactoring the source code as a means to not unintentionally introduce bugs~\cite{40}. Furthermore, developers feel discouraged from applying refactorings due to the challenge of maintaining backward compatibility~\cite{40}. 
Moreover, one could expect that when developers apply root-canal refactoring, the refactored code element is less susceptible to containing bugs than when developers apply floss refactoring. In the example, the developer performed a floss refactoring that originated a bug. Despite refactoring the code, the developer was concerned about maintaining the backward compatibility of Spring Boot. Thus, the developers might have been overloaded by this floss refactoring, and inadvertently introduced such a bug.

Unfortunately, there is limited understanding about the bug-proneness of refactored code, taking into account the various characteristics of refactoring. Existing studies~\cite{1, 3} provide evidence that refactored code is often susceptible to contain bugs. However, none of the existing studies~\cite{1, 3} analyze how floss refactoring compare to root-canal refactorings regarding bug proneness. For instance, in the example above, the floss refactoring performed by the developer could have made the code element bug-prone if compared to when the developer applies a root-canal refactoring. Furthermore, some refactoring types might be more complex than others, therefore, increasing the susceptibility of bugs in the refactored code. The lack of knowledge about the characteristics of refactorings that make code elements bug-prone prevents researchers and practitioners from having better practices when applying refactorings. The general problem is that the characteristics of refactorings that make refactored code elements bug-prone remain unknown.




%

\section{Related Work}
\label{subsec:related-work}

Only a few studies have investigated the bug-proneness of refactored code. Bavota et al.~\cite{1} analyzed whether refactored code elements contain a bug in further commits. The authors focus on analyzing: (i) the percentage of refactored code elements that become buggy and (ii) the refactoring types that are more likely to make the refactored code bug-prone. Their results show that refactored code is often bug-prone. They also found that some refactoring types are more harmful than others, such as refactorings involving hierarchies (e.g., Pull Up Method). This is also evident in recent work by Counsel et al.~\cite{69} where they found that the Change Variable refactoring accounted for 80\% of buggy code refactorings. However, these studies do not address how close a refactored code is to becoming bug-prone, but rather the percentage of refactorings that resulted in buggy code. The authors only take into consideration if a refactored code contains at least one bug in further commits. This analysis might not suffice to measure the bug-proneness of refactored code because a bug might have emerged in the same code element many commits away from the refactoring. Also, the refactoring data were not manually analyzed to confirm whether the refactoring indeed made the source code bug-prone.  

Di Penta et al.~\cite{67} examined the relationship between refactoring and bug-fixing activities by employing both qualitative and quantitative analyses of refactorings and associated bugs. Their findings indicate that 38\% of refactoring actions resulted in subsequent bug-fixing activities. A study conducted by Amirreza and Péter~\cite{68} using the SmartSHARK dataset suggests that 54\% of bug-inducing commits are associated with refactorings. 
These two studies do not address the level of proneness or the proximity of refactored code to potential bugs, and also, their approach excluded refactored code linked to reported but unresolved bugs, thereby losing potentially valuable insights.

Conversely, Wei{\ss}gerber and Diehl~\cite{3} investigate if the number of bug reports opened in the next five days after the refactoring increases or decreases. That is, the authors investigate whether the bug-proneness of refactored code increases or decreases according to the number of refactorings and opened bug reports. They found that a high ratio of refactorings is often followed by an increasing ratio of bug reports. The downside of this study is that the refactored code may not be executed to exhibit its associated bug within the next five days (as in our motivating example in Section~\ref{sec:motivating-example}), and hence, future bug incidents may not be accounted for. Thus, for changes between two major releases, many refactorings and bugs are hidden or unidentifiable in their analyses.

In addition to the limitations presented above, there is no study that investigates the various refactoring characteristics together with bug-proneness. For instance, the authors overlook the impact of different refactoring purposes (root-canal vs. floss) and the number of refactorings (single vs. composite) on the bug-proneness of refactored code. Given all these limitations, we propose a different approach for evaluating the bug-proneness of refactored code, as discussed next.


\section{Study Design}
\label{sec:methodology}

The goal of this study is to \textit{analyze the bug-proneness of refactored code according to various refactoring characteristics in the context of code smells, addressing the limitations of the literature}. In the following section, we present the research questions (RQs) and the steps of our study.

\subsection{Research Questions}

Our study is guided by five RQs, as follows.



\vspace{1mm} \noindent \textit{\textbf{RQ$_1$. Is refactored code less or more bug-prone than non-refactored code?}}
This RQ serves as a starting point for understanding the extent to which we may attribute the introduction of bugs during software maintenance to refactorings. Since developers often apply refactorings on the degraded code elements~\cite{6}, in this RQ we analyze code elements that contain code smells before the refactoring operation. Then, we compute the number of code elements that were refactored and later became buggy or not.

\vspace{1mm} \noindent
\textit{\textbf{RQ$_2$. How frequent are code elements bug-prone per refactoring type?}} 
There are various types of refactorings applied by developers~\cite{ivers2022industry,szHoke2017empirical,alomar2021refactoring}. These refactorings target different structures of the code~\cite{19}, such as classes (e.g., Move Class), methods (e.g., Extract Method and Inline Method), element names (e.g., Rename Class and Rename Method), or even the classes hierarchies (e.g., Pull Up Method and Push Down Method). Given the different complexity of each refactoring, we seek to know how often each refactoring type leads to code becoming buggy. Gaining this knowledge enables us to guide developers in making safer refactoring choices, ultimately enhancing software maintenance. 

\vspace{1mm} \noindent
\textit{\textbf{RQ$_3$. What is the bug-proneness of code elements that underwent single and composite refactorings?}}
Refactoring operations can be applied individually (single refactoring), e.g., by applying a method rename; or in a composite manner, such as applying an Extract Method followed by a Method Rename~\cite{Murphy-Hill2012}. It is known that sometimes single refactoring is not enough to remove code smells~\cite{cedrim2017understanding, bibiano2019quantitative,bibiano2023composite}, and also that composite refactorings can lead to side effects~\cite{bibiano2021look}. To answer RQ$_3$, we analyzed how many code elements underwent single and composite refactorings and later became buggy. 




\vspace{1mm} \noindent
\textit{\textbf{RQ$_4$. What is the impact of refactoring on bug-proneness when comparing different refactoring purposes?}}
As refactorings are applied with different intents, namely either to improve the code
structure quality, or in conjunction with adding new features (similar to the motivating example in Section~\ref{sec:motivating-example}) or fixing bugs~\cite{paixao2020behind}. Thus, in this RQ we seek to know the impact of floss or root-canal refactorings on the bug-proneness of the refactored code. 


\vspace{1mm} \noindent
\textit{\textbf{RQ$_5$. How long does it take for a refactored code element to become buggy?}} 
For this question, we define the notion of distance. \textit{Distance} is the number of commits (or changes) between the commit in which developers have applied the refactoring and the commit in which the bug emerged. A bug can emerge either from the commit in which the bug was indeed introduced in the code element or the commit related to when a developer or a user opened a bug report. We consider both events of bugs. We can say that the smaller the number of changes or commits between the refactoring and the bug (shorter distance), the higher the bug-proneness of refactored code elements.
Thus, for this RQ, we seek to measure how many further changes are required for either a single or composite refactored code to become buggy, and the duration it takes for it to occur. We compute the distance properties for each of the single and composite refactorings and analyze the time frame for which a bug report was raised. This analysis can shed light on the stability of refactored code and inform better refactoring practices to reduce the bug-proneness of refactored code.




\subsection{Study steps}

To reach our goal and answer our RQs, we performed a longitudinal multi-project study assessing the bug-proneness of code elements in the light of various refactoring characteristics. We present a description of our study steps as follows.

\vspace{1mm} \noindent \textit{\textbf{Project Selection.}}
Our studies involved 12 popular Java-based open source projects with active usage of their issue tracking systems, such as Bugzilla and the GitHub issue management system, for bug reporting. This is due to the wide adoption of Java\footnote{\url{https://www.tiobe.com/tiobe-index/}} and the fact that open source projects will ease further replication. While selecting these projects, we also considered projects in which developers have the culture of describing the bug report being fixed in their commit messages. 

Table~\ref{tab:selected-projects} presents details of the projects selected for our study. 
The dataset includes shorter periods, like two years and eight months for \textit{Fresco}, capturing potentially more focused efforts. We also have projects with longer time spans, up to 16 years and six months such as \textit{Ant}, likely reflecting extensive code evolution. This variability in duration highlights diverse project lifecycles, with an average duration of six years and 10 months across all projects, allowing for comprehensive refactoring insights across different development phases.

\begin{table}
\centering
\small
\caption{General data of the analyzed software projects}
\begin{tabular}{lcr}
\toprule
\textbf{Project} & \textbf{Analyzed Period} & \textbf{No. Commits}   \\
\midrule
Ant & 2000-01 to 2016-07& 13,331\\
Derby& 2004-08 to 2017-08&7,865 \\
Elasticsearch& 2010-08 to 2016-08 & 23,597 \\
Elasticsearch-hadoop& 2013-11 to 2017-11 & 1,718 \\
ExoPlayer& 2014-06 to 2017-10 & 3,081 \\
Fresco& 2015-03 to 2017-11 & 1,535 \\
Material-dialogs& 2014-05 to 2017-10 & 1,330 \\
Netty& 2008-08 to 2017-11 & 8,357 \\
OkHttp& 2011-05 to 2016-08 & 2,645 \\
Presto& 2012-08 to 2016-08 & 8,056 \\
Spring-boot& 2012-10 to 2016-08 & 8,529 \\
Tomcat& 2006-03 to 2016-12 & 17,732 \\
\midrule
\end{tabular}
\label{tab:selected-projects}
\end{table}

\vspace{1mm} \noindent \textit{\textbf{Code Smell Detection.}}
To assess the bug-proneness of refactored in the context of different refactoring characteristics, we focus on the analysis of degraded code elements, namely pieces of code affected by code smells~\cite{19,oliveira2022developers}. This focus is because: (i) developers apply refactoring with different purposes (i.e., root-canal and floss refactorings) targeting restructuring degraded code elements, and (ii) developers much more often choose to apply refactorings in degraded code elements~\cite{6, 32}.

Code smells are often identified with rule-based strategies~\cite{15}. Each of these strategies requires the computation of certain metrics and comparison with their thresholds. In this context, we selected specific rules, metrics and thresholds from previous works to support the identification of code smells~\cite{16, 17}. These rules were chosen because they are refinements of the well-known rules and have a precision of 72\% and recall of 81\% on average~\cite{17}. In total, we analyzed 17 types of code smells. Further details on these are available in our supplementary material\cite{supplementary}. We selected these code smells because they are very common and are directly related to the most frequent refactoring types~\cite{19, 21, 31}.

\vspace{1mm} \noindent \textit{\textbf{Refactoring Detection and Manual Validation.}}
We used the Refactoring Miner tool to identify refactorings in selected projects due to its precision of 98\% and a recall of 87\% as stated by Tsantalis et al.~\cite{20, 66}. We choose to study the 13 most commonly investigated refactoring types in the literature~\cite{18}. A complete list of analyzed refactoring types is available in our supplementary material for this paper~\cite{supplementary}. These refactoring types are defined in the Fowler’s catalog~\cite{19}. We collected 27,450 refactoring instances in total. More specifically, we found 19,622 single refactorings and 7,828 composite refactorings. We classified refactorings as composites if (i) the refactorings have at least one code element in common~\cite{18, 40}; (ii) all refactorings in the composite must have been performed by the same developer; and (iii) the change must have more than one refactoring operation.

To ensure the reliability of our data, we conducted a manual validation of the refactorings identified by the Refactoring Miner tool. This validation covered a random set of 2,119 refactoring instances from different refactoring types, since the precision of the Refactoring Miner tool could vary due to the rules implemented to detect each refactoring type. These 2,119 refactorings were manually validated for the refactoring purpose (i.e., root-canal vs. floss).


\vspace{1mm} \noindent \textit{\textbf{Bug Detection and Manual Validation.}}
We identified bugs for the selected projects either from their \textit{bug reports} or \textit{via static analysis tools}.
Bugs from bug reports are found by users or developers during production usage or testing. For these, we selected bug reports with the status resolved fixed, verified fixed, closed, closed fixed, labeled as ``bug'' or ``defect'' in the analyzed issue tracking systems. We collected 6,051 bug reports, including the commit on the date before the bug report was opened. The projects \textit{Apache Tomcat}, \textit{Apache Ant}, and \textit{Apache Derby} use Bugzilla\footnote{\url{https://www.bugzilla.org/}} as the issue tracking system for bug reports. The projects \textit{OkHttp} and \textit{Prestouse} use GitHub\footnote{\url{https://docs.github.com/en/issues/tracking-your-work-with-issues}} as issue tracking systems.
Bugs were also identified via FindBugs, version 3.0.1~\cite{51}, which is a static analysis tool. FindBugs detects dormant bugs in the source code that have not been reported or identified yet for each commit. We identified 49,250 with this tool for the projects \textit{Apache Derby} and \textit{Apache Tomcat}. Table~\ref{tab:bug-patterns-findbug} presents the description of bug patterns selected for our study.

\begin{table}
\centering
\small
\renewcommand{\arraystretch}{1.5}
\caption{Bug Patterns of FindBugs}
\begin{tabular}{m{2cm} m{6cm}}
\toprule
\textbf{Bug Pattern} & \textbf{Description}    \\
\midrule
Malicious Code & Variables or fields exposed to classes that should not be using them \\
Multithreaded & Thread synchronization issues \\
Performance & Similar to malicious code vulnerability \\
Security & How frequent degraded code elements are bug-prone per refactoring type?                  \\
Correctness & Apparent coding mistakes \\
\bottomrule
\end{tabular}
\label{tab:bug-patterns-findbug}
\end{table}

As mentioned in previous research that bug report classification are unreliable~\cite{23} we performed manual classification of bug reports to identify which bug reports actually represent bugs. This was done in pairs by 14 researchers with each person of the pair responsible for manually classify the same bug report as ``bug'' or ``not bug''. When there was a divergence in opinion, the pair discussed final classification of such bug. We manually validated 1,477 bug reports of which 516 (34.94\%) were classified as ``bug'' and 961 (65.06\%) as ``not bug''. We also performed a manual validation of the bugs collected via static analysis in order to reduce the number of false positives of the tool. To do that, we followed the same procedure described above for the bug reports. We manually validated 198 bugs of which 168 (84.85\%) were classified as ``bug'' and 30 (15.15\%) as ``not bug''. In total, we collected 6,051 bug reports and 49,250 bugs via static analysis tools.

\vspace{1mm} \noindent \textit{\textbf{Bug-Fix Commit and Fixed Code Element Detection.}}
In order to identify the bug-inducing commit, it was necessary for us to know the bug-fix commit, and the code elements involved with the fix of the bug. We relied on the common practices where developers include the bug report number in the commit comment whenever they fix a bug associated with it~\cite{24}. In this way, to map a bug report with its fix commit, we automatically search log messages for references to bug reports such as ``bug 23442'' or ``fix for bug 23442'' as proposed by Dallmeier and Zimmermann~\cite{25}.
This process was necessary for bug reports only, since the bugs detected via static analysis tools provide their commits and code elements associated with their fixes.
We ignored bug reports for which we could not find commits of their fixes because, without the commits, we could not find the fixed files. Thus, we classified these bug reports as not functional~\cite{26}. 

\vspace{1mm} \noindent \textit{\textbf{Bug-Introducing Commit Detection.}}
Given the bug-fix commit and the bug-fix elements previously identified, we used the SZZ algorithm to identify when the bug was introduced in the project, since it is the most used algorithm for such purposes~\cite{27}. SZZ is a bug-introducing change identification algorithm proposed by Śliwerski et al.~\cite{24}. However, the original SZZ algorithm has some limitations, as mentioned by Kim et al.~\cite{28} and Williams and Spacco~\cite{20}. The first is that not all changes are bug fixes because even if a file change is defined as a bug fix by developers, not all hunks in the change are bug fixes. The second limitation is due to the insufficient information in bug tracking systems, which may lead to incorrect choosing of bug-inducing commits. For these reasons, we used a combination of heuristics proposed by Kim et al. and Williams et al. by using their approach, we can remove 38- 51\% of false positives and 14\% of false negatives as compared to the original implementation of SZZ~\cite{24}. The algorithm outputs a list of commits related to the introduction of the bug in the software system. For analysis purposes, we considered only the newest commit reported by SZZ.

\subsection{Measuring the bug-proneness of refactored code elements}

\begin{figure}[!tp]
    \centering
    \includegraphics[width=1\linewidth]{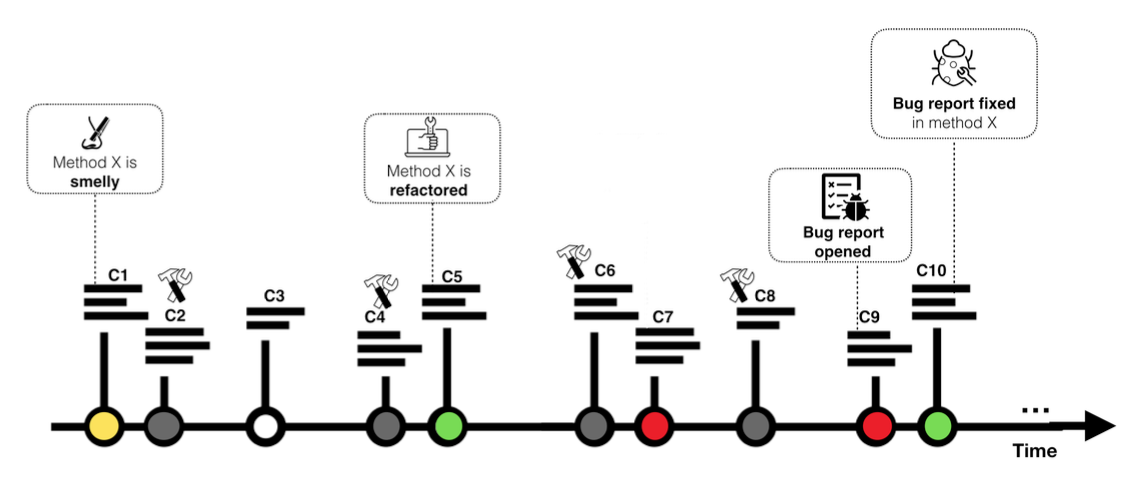}
    \caption{Evaluation of the bug-proneness of refactored code}
    \label{fig:commits-timeline}
\end{figure}

Figure~\ref{fig:commits-timeline} illustrates how we evaluate the bug-proneness of refactored code for a given refactoring (either single or composite). For simplicity, commits are presented as timelines for events performed in the same code element (i.e., method X in this figure). We can see method X was smelly in the first commit C1, after which two changes were performed in the commits C2 and C4. The commit C3 was performed with other operations not relevant for method X. In C5, a refactoring of any type is applied, which was followed by another change in C6. Then, a bug was detected and reported by users in the commit C9. Between the bug introduction and the bug report, another change was performed in method X, namely C8. Finally, the bug was fixed for method X in C10.

To measure the distance between a refactoring and a code become buggy, we count the number of changes between the commit where the refactoring was applied and the commit where the bug was either introduced or reported. This is done by examining all code elements involved in the bug fix. For instance, in Figure~\ref{fig:commits-timeline}, using C5 as a reference point, the distance is 1 at C7 (where the bug was introduced) and 2 at C9 (where the bug was reported).

The frequency property is obtained by counting the number of code refactorings that match the condition of being studied (e.g., refactoring type, floss vs. root-canal, single vs. composite). For instance, to answer RQ$_2$, we count the number of refactorings that became buggy and group them by the available refactoring types. Similarly, to answer RQ$_3$, we analyze and count separately the number of single and composite refactorings that later became buggy. 




\section{Results And Discussion} 
This section presents the results of our findings from the analysis of 27,450 refactorings from the 12 open-source Java-based projects. Also, we provide answers to our research questions discussed in Section~\ref{sec:methodology}.

\begin{table*}
\caption{Frequency of the bug-proneness of refactored code vs. non-refactored code, and the Proportion of Buggy Refactoring, for both single and composite refactorings}
\label{tab:bug_proness_frequency}
\centering
\small
\begin{tabular}{l|rr|rr|r} 
\hline
\multirow{3}{*}{\textbf{Project}} & \multicolumn{2}{c|}{\textbf{No Refactoring}} & \multicolumn{2}{c|}{\textbf{Refactoring}} & \multirow{3}{*}{\textbf{Proportion}} \\ \cline{2-5}
& \textbf{\makecell{Single smell\\ \& Buggy} }& \textbf{\makecell{Multiple smells\\ \& Buggy}} & \textbf{\makecell{Single smell\\ \& Buggy}} & \textbf{\makecell{Multiple smells\\ \& Buggy}} & \\
\hline
\hline
\multicolumn{6}{c}{\textbf{Single Refactoring}}\\
\hline
\hline
Ant               & \cellcolor{cellh}29.63\% & \cellcolor{cellh}37.04\% & 0.00\%  & \cellcolor{cellh}33.33\%  & \cellcolor{cellh}33.33\%  \\
Derby             & 7.52\%  & \cellcolor{cellh}91.23\% & 0.00\%  & 1.25\%   & 1.25\%   \\
Elasticsearch     & \cellcolor{cellh}39.95\% & \cellcolor{cellh}31.09\% & 2.55\%  & \cellcolor{cellh}26.42\%  & \cellcolor{cellh}28.97\%  \\
Elasticsearch-hadoop & 0.00\%  & 0.00\%  & 0.00\%  & 0.00\%   & 0.00\%       \\
ExoPlayer         & \cellcolor{cellh}71.43\% & 0.00\%  & \cellcolor{cellh}28.57\% & 0.00\%   & \cellcolor{cellh}28.57\%  \\
Fresco            & \cellcolor{cellh}82.64\% & \cellcolor{cellh}17.36\% & 0.00\%  & 0.00\%   & 0.00\%   \\
Material-dialogs  & 0.00\%  & 1.03\%  & 0.00\%  & \cellcolor{cellh}98.97\%  & \cellcolor{cellh}98.97\%\\
Netty             & \cellcolor{cellh}60.26\% & \cellcolor{cellh}38.46\% & 1.28\%  & 0.00\%   & 1.28\%   \\
Okhttp            & 7.69\%  & \cellcolor{cellh}80.77\% & 0.00\%  & \cellcolor{cellh}11.54\%  & \cellcolor{cellh}11.54\%  \\
Presto            & \cellcolor{cellh}32.08\% & \cellcolor{cellh}39.62\% & 7.55\%  & \cellcolor{cellh}20.75\%  & \cellcolor{cellh}28.30\%  \\
Spring-boot       & \cellcolor{cellh}17.18\% & \cellcolor{cellh}30.53\% & 3.05\%  & \cellcolor{cellh}49.24\%  & \cellcolor{cellh}52.99\% \\
Tomcat            & \cellcolor{cellh}74.84\% & \cellcolor{cellh}24.74\% & 0.00\%  & 0.42\%   & 0.42\%   
\\

\hline
\hline
\multicolumn{6}{c}{\textbf{Composite Refactorings}}\\
\hline
\hline
Ant & 6.25\% & \cellcolor{cellh}71.88\% & 0.00\% & \cellcolor{cellh}21.88\% & \cellcolor{cellh}21.88\% \\ 
Derby & \cellcolor{cellh}14.29\% & \cellcolor{cellh}84.07\% & 0.00\% & 1.65\% & 1.68\% \\ 
Elasticsearch & 0.00\% & 0.00\% & 0.00\% & 0.00\% & 0.00\% \\ 
Elasticsearch-hadoop & 0.00\% & 0.00\% & 0.00\% & 0.00\% & 0.00\% \\ 
ExoPlayer & 0.00\% & 0.00\% & 0.00\% & 0.00\% & 0.00\% \\ 
Fresco & 0.00\% & \cellcolor{cellh}98.29\% & 1.71\% & 0.00\% & 1.71\% \\ 
Material-dialogs & 0.00\% & \cellcolor{cellh}100.00\% & 0.00\% & 0.00\% & 0.00\% \\ 
Netty & \cellcolor{cellh}49.30\% & \cellcolor{cellh}11.27\% & 2.82\% & \cellcolor{cellh}36.62\% & \cellcolor{cellh}39.44\% \\ 
Okhttp & \cellcolor{cellh}11.76\% & \cellcolor{cellh}64.71\% & 0.00\% & \cellcolor{cellh}23.53\% & \cellcolor{cellh}23.53\% \\ 
Presto & \cellcolor{cellh}37.50\% & \cellcolor{cellh}52.50\% & 0.00\% & \cellcolor{cellh}10.00\% & \cellcolor{cellh}10.00\% \\ 
Spring-boot & \cellcolor{cellh}11.92\% & \cellcolor{cellh}43.05\% & 0.00\% & \cellcolor{cellh}45.03\% & \cellcolor{cellh}45.03\% \\ 
Tomcat & \cellcolor{cellh}75.00\% & \cellcolor{cellh}24.56\% & 0.00\% & 0.44\% & 0.44\% \\ \hline
\end{tabular}
\end{table*}

\subsection{RQ$_1$. Is refactored code less or more bug-prone than non-refactored code?}
Previous work mentions that refactored code elements are often noticed to have a few bugs reported~\cite{3,67}. To check this phenomenon in our dataset, for single and composite refactorings, we computed the number of buggy elements that had either been refactored or not refactored and further narrowed the computation to either the presence of one or multiple code smells. Additionally, to corroborate our analysis, we computed the proportion (P) of the code elements with single or multiple smells that were refactored and became buggy to the total smelly refactored and non-refactored code. This is shown in the Equation~\ref{eq:proportion}.
\begin{equation}
    \small
    P = \frac{\#\left(\parbox{5cm}{\centering buggy refactored code with \\ single or multiple smells}\right)}{\#\left(\parbox{6cm}{\centering buggy refactored and non-refactored code \\ with single or multiple smells}\right)}
    \label{eq:proportion}
\end{equation}


Table~\ref{tab:bug_proness_frequency} presents the frequency of buggy elements found in refactored and non-refactored code based on the number of code smells (i.e., single code smell or multiple code smells) for single and composite refactorings. 
At first, we see that the systems have different adoption of single and composite refactorings. Also, there is no pattern in the number of smells and refactorings being the reason for making a code more or less bug-prone. In our dataset, no instances of smelly or single refactorings led to buggy code in \textit{Elasticsearch-Hadoop}. Similarly, for composite refactorings across \textit{Elasticsearch}, \textit{Elasticsearch-Hadoop}, and \textit{ExoPlayer}, we observed no occurrences of smelly or composite refactored code elements becoming buggy, as indicated as 0\% frequency in Table~\ref{tab:bug_proness_frequency}.
Focusing on the proportion of refactorings that make a code buggy (last column in Table~\ref{tab:bug_proness_frequency}),  we saw that \textit{Material-dialog} and \textit{Spring-boot} contained a high proportion, respectively 98.97\% and 52.99\%, of smelly code that was refactored with single refactorings and later found to be buggy. Other projects like \textit{Ant}, \textit{Elasticsearch-hadoop}, \textit{ExoPlayer}, and \textit{Presto} had quite a reasonable proportion of their smelly and single refactored code become buggy.
For composite refactorings, the greatest values for the proportion are for the systems \textit{Spring-boot} and \textit{Netty}, respectively 45.03\% and 39.44\%. Also, \textit{Okhttp} and \textit{Ant} had more than 20\% smelly code that was refactored with composite refactorings and later found to be buggy.


 \begin{figure}[!tp]
    \centering
    \includegraphics[width=0.9\linewidth]{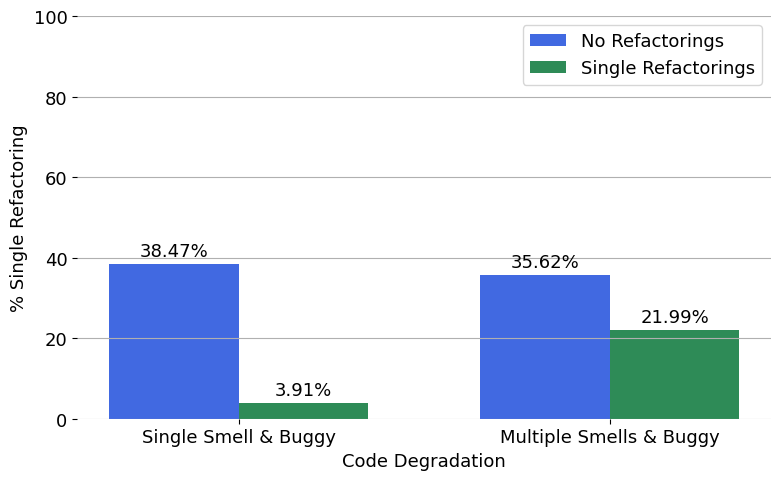}
    \caption{Frequency of single refactorings according to the structure degradation of the code elements touched by these refactorings}
    \label{fig:single_freq_bug_prone_ref_vs_non_refac}
\end{figure}
\begin{figure}[!tp]
    \centering
    \includegraphics[width=0.9\linewidth]{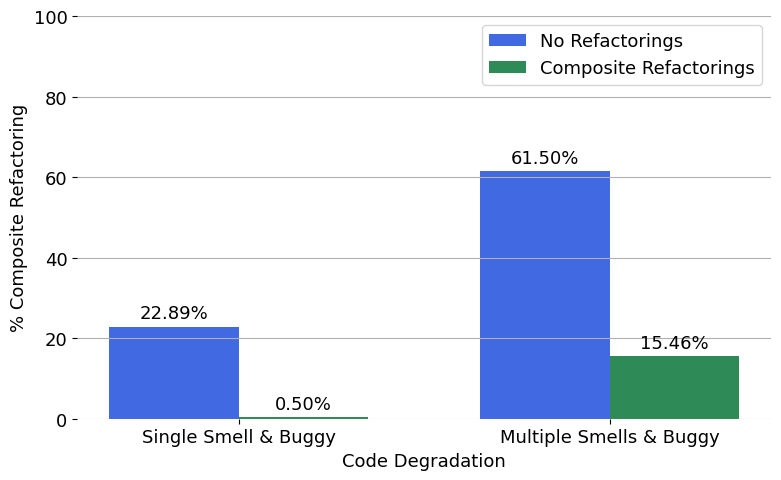}
    \caption{Frequency of the bug-proneness of refactored code vs. non-refactored code (composite refactorings)}
    \label{fig:composite_bug_prones_refac_vs_non_refac}
\end{figure}



An overall summary of the data for all the 12 projects according to the number of code smells affecting the code elements before single refactoring is presented in Figures~\ref{fig:single_freq_bug_prone_ref_vs_non_refac} and~\ref{fig:composite_bug_prones_refac_vs_non_refac}, for single refactorings and composite refactorings, respectively. In the context of single refactoring, we see in Figure~\ref{fig:single_freq_bug_prone_ref_vs_non_refac} that 57.61\% of the buggy code elements contain multiple smells (i.e., sum of blue and green bars for multiple smells, on the right side of the figure) against 42.38\% that contain single smells (i.e., sum of blue and green bars for single smell, on the left side of the figure). Thus, we observe that, for the projects under analysis, code elements with multiple smells are more bug-prone than code elements with single smells, irrespective of whether the code has been refactored. 
Similarly, in Figure~\ref{fig:composite_bug_prones_refac_vs_non_refac}, we can see that 76.61\% of the buggy code elements contain multiple smells, against 23.39\% that contain single smells regardless of being composite refactored or not. Yet, for single refactoring, we can also say that code elements with multiple smells are more bug-prone than code elements with single smells, irrespective of whether the code has been refactored in a composite manner. Thus, we state that the presence of multiple smells in code elements makes the code more bug-prone than the presence of single smells. 



We also analyzed the frequency of non-refactored code that became buggy. In Figure~\ref{fig:single_freq_bug_prone_ref_vs_non_refac} we can see that this happens in 74.09\% of the cases (sum of blue bars), while the refactored code elements that became buggy is 25.9\% (sum of green bars). For composite refactoring (Figure~\ref{fig:composite_bug_prones_refac_vs_non_refac}), the frequency of non-refactored code that became buggy is 84.04\% (sum of blue bars), while that of refactored code elements that became buggy is 15.96\% (sum of green bars).

To check the statistical significance of our results, we applied the Fisher's test to compute the strength of the relation between composite refactoring a degraded code with bugs~\cite{9}. Furthermore, we used the Odds Ratio~\cite{10} to check the possibility of the presence or absence of a phenomenon (i.e., composite refactoring) to be associated with the presence or absence of the other phenomenon (i.e., degraded code with bugs). Considering all projects analyzed, we found a p-value less than 0.05, and the Odds Ratio equals 0.2894 and 0.6881 for single and composite refactorings, respectively. Thus, our results show that the possibility of composite refactoring being related to a degraded code with bugs is 0.68 if compared to non-refactoring related to degraded code with bugs while that of single refactoring is 0.28.


\begin{tcolorbox}[colframe=gray,colback=white,boxrule=2pt,arc=.3em,boxsep=-1mm]
\textbf{RQ$_1$ Answer:} Our results show that refactored code elements are less bug-prone than non-refactored code, since a high proportion of the refactored code did not become buggy even though they contained at least one code smell. This confirms the findings of previous work. Wei{\ss} Gerber and Diehl~\cite{3} observed a prevailing number of refactoring phases followed by a few reported bugs. Di Penta et al. found that in 62\% of the cases they analyzed, they could not find evidence for bug fixes for the refactored code~\cite{67}.
\end{tcolorbox}

\subsection{RQ$_2$. How frequent are code elements bug-prone per refactoring type?}

\begin{table}[!t]
\centering
\caption{Frequency of Bugs per refactoring type}
\begin{tabular}{l c l}
\hline
\textbf{Refactoring Type} & \textbf{Frequency} & \\
\hline
Extract Method      & 54.00\% & \rule{2.7cm}{6pt} \\
Inline Method       & 25.00\% & \rule{1.3cm}{6pt} \\
Rename Method       & 7.00\%  & \rule{0.5cm}{6pt} \\
Move Attribute      & 6.00\%  & \rule{0.4cm}{6pt} \\
Move Method         & 5.00\%  & \rule{0.3cm}{6pt} \\
Extract Superclass  & 3.00\%  & \rule{0.1cm}{6pt} \\
\hline
\label{tab:single_refactorings_per_type}
\end{tabular}
\end{table}

To answer RQ$_2$, we analyze which of the 13 refactoring types considered in our study are often related to the occurrence of bugs during single refactoring. As shown in Table~\ref{tab:single_refactorings_per_type}, we found that the most commonly used refactoring technique was the Extract Method, accounting for 54\%, followed by the Inline Method at 25\%, the Rename Method at 7\%, Move Attribute at 6\%, the Move Method at 5\%, and Extract Superclass at 3\%. The remaining refactoring types did not have any value of frequency. 
We found that code elements refactored by Extract Interface, Move Class, Pull up Attribute, Pull down Attribute, Pull up Method, Push down Method, and Rename Class have no relation to bug proneness in our dataset. 

\begin{tcolorbox}[colframe=gray,colback=white,boxrule=2pt,arc=.3em,boxsep=-1mm]
\textbf{RQ$_2$ Answer:} Bugs were more frequently found in code refactored by the Extract Method and Inline Method, which confirms the claims by previous work~\cite{1} that these refactoring types are very harmful. Hence, developers should be aware when applying Extract Method and Inline Method. Surprisingly, this result contradicts Bavota et al.~\cite{1}, which found that those refactoring types are one of the most harmful, especially refactoring types involving hierarchies (e.g., Pull up Method, and Pull down Method).
\end{tcolorbox} 


\subsection{RQ$_3$. What is the bug-proneness of code elements that underwent single and composite refactorings?}
\begin{table*}[h!]
\centering
\small
\caption{Frequency of single and composite refactorings performed in code elements that are bug-prone per software project}
\label{tab:merged_refactorings}
\begin{tabular}{m{2.5cm}| r r  r| r r r}
\hline
\multirow{2}{*}{\textbf{Project}} & 
\multicolumn{3}{c|}{\textbf{Single Refactorings}} & 
\multicolumn{3}{c}{\textbf{Composite Refactorings}} \\ \cline{2-7}
 & \textbf{No. Refactorings} & \textbf{Bug-prone} & \textbf{Not Bug-prone} & \textbf{No. Refactorings} & \textbf{Bug-prone} & \textbf{Not Bug-prone} \\ \hline
Ant                  & 1,276    & 11 (0.86\%)      & 1,265 (99.14\%)     & 331    & 2 (0.60\%)       & 329 (99.40\%)       \\ 
Derby                & 3,093    & 22 (0.71\%)      & 3,071 (99.29\%)     & 4,407  & 3 (0.07\%)       & 4,404 (99.93\%)     \\ 
Elasticsearch        & 5,597    & 736 (13.15\%)    & 4,861 (86.85\%)     & 816    & 72 (8.82\%)      & 744 (91.18\%)       \\ 
Elasticsearch-hadoop & 265      & 3 (1.13\%)       & 262 (98.87\%)       & 242    & 0 (0.00\%)       & 242 (100.00\%)      \\ 
ExoPlayer            & 1,198    & 21 (1.75\%)      & 1,177 (98.25\%)     & 410    & 6 (1.46\%)       & 404 (98.54\%)       \\ 
Fresco               & 564      & 5 (0.89\%)       & 559 (99.11\%)       & 18     & 1 (5.56\%)       & 17 (94.44\%)        \\ 
Material-dialogs     & 78       & 12 (15.38\%)     & 66 (84.62\%)        & 18     & 6 (33.33\%)      & 12 (66.67\%)        \\ 
Netty                & 2,782    & 31 (1.11\%)      & 2,751 (98.89\%)     & 766    & 12 (1.57\%)      & 754 (98.43\%)       \\ 
Okhttp               & 698      & 47 (6.73\%)      & 651 (93.27\%)       & 66     & 5 (7.58\%)       & 61 (92.42\%)        \\ 
Presto               & 1,526    & 32 (2.10\%)      & 1,494 (97.90\%)     & 250    & 4 (1.60\%)       & 246 (98.40\%)       \\ 
Spring-boot          & 1,152    & 176 (15.28\%)    & 976 (84.72\%)       & 130    & 21 (16.15\%)     & 109 (83.85\%)       \\ 
Tomcat               & 1,393    & 36 (2.58\%)      & 1,357 (97.42\%)     & 374    & 5 (1.34\%)       & 369 (98.66\%)       \\ \hline
\textbf{Total}       & \textbf{19,622} & \textbf{1,132 (5.77\%)} & \textbf{18,490 (94.23\%)} & \textbf{7,828} & \textbf{137 (1.75\%)} & \textbf{7,691 (98.25\%)} \\ \hline
\end{tabular}
\end{table*}

Table~\ref{tab:merged_refactorings} presents the results on single and composite refactorings. The table also shows a breakdown of the refactored code was later found to become buggy or not. We analyzed 19,622 single refactorings and 7,828 composite refactorings, as shown in the last row of the table. The projects \textit{Spring-boot}, \textit{Elasticsearch}, and \textit{Material-dialogs} each had more than 10\% of their refactorings performed are bug-prone, while for the other projects varied from 0.07\% to 8.82\%.
 
Overall, the results in Table~\ref{tab:merged_refactorings} show that only 5.77\%  of the single refactorings were performed in code elements that became buggy, against 94.23\% of single refactorings that were performed in code elements that did not become buggy. For composite refactorings, the table shows that 1.75\% of them were later found to become buggy, while the remaining 98.25\% did not become buggy. Thus, single refactoring is more than three times more bug prone than composite refactorings in our dataset. 
The potential explanation is that single refactoring are commonly applied with other code changes (floss refactoring)~\cite{19,Murphy-Hill2012}, which can increase the bug proneness of refactored code~\cite{paixao2020behind}. 


We also compare the proportion of smelly code that was refactored with single refactorings and that became buggy, to the smelly code refactored with composite refactoring. In Table~\ref{tab:merged_refactorings}, we can see that in the majority of the projects, the proportions of smelly and single refactored code that became buggy is higher than composite refactoring. For instance, \textit{Spring-boot} has 52.99\% proportions of its smelly code that were refactored in isolation while that of its composite refactoring is 45\%. Similar situation is observed for \textit{Ant} and \textit{Elasticsearch}. These results are interesting because they indicate that smelly code can be bug-prone when they are not fully refactored. Studies indicate that single refactorings often do not fully remove code smells~\cite{cedrim2017understanding, bibiano2019quantitative}. Thus, the code smells can remain after the application of single refactorings, increasing the degradation of internal software quality. This degradation when combined with other code changes can increase the occurrences of bugs. In addition, recent studies~\cite{bibiano2020how, bibiano2021look, bibiano2024enhancing} demonstrate that composite refactorings tend to keep the internal software quality, regardless of the full code smell removal. This indicates that even when other code changes are applied, the bug-prone is reduced when the internal software quality is kept after composite refactorings.


\begin{tcolorbox}[colframe=gray,colback=white,boxrule=2pt,arc=.3em,boxsep=-1mm]
\textbf{RQ$_3$ Answer:} When comparing single and composite refactoring, the frequency of bug occurrence for bug occurrence of single refactoring is higher than composite refactoring. Hence, single refactoring is more bug-prone than composite refactoring in terms of frequency of bugs.
\end{tcolorbox}


\subsection{RQ$_4$. What is the impact of refactoring on bug-proneness when comparing different refactoring purposes?}

\begin{table}[!tp]
\centering
\caption{Summary of Distance of Bug insertion and Report events for Single Refactoring}
\begin{tabular}{llrrrrrr}
\hline
\textbf{Purpose} & \textbf{Bug}    & \textbf{N} & \textbf{Min} & \textbf{Q1} & \textbf{Q2} & \textbf{Q3} & \textbf{Max} \\
\hline
\multirow{2}{*}{Root-canal} & Insertion & 23  & 1   & 11.2 & 20  & 27  & 61  \\
& Report    & 74  & 0   & 7    & 19  & 42.1 & 87  \\
\hline
\multirow{2}{*}{Floss} & Insertion & 723  & 1   & 7    & 16.5 & 32  & 271 \\
& Report    & 2203 & 0   & 8    & 19   & 38  & 364 \\
\hline
\end{tabular}
\label{tab:root-canal-floss}
\end{table}

Table~\ref{tab:root-canal-floss} presents the results for the analysis of bug proneness of degraded code refactored with root-canal and floss refactorings. The columns Min, Q1, Q2, and Q3, and Max represent the minimum value of distance, the quartiles (25\%, 50\%, and 75\%), and the maximum distance, respectively.

When we compare the distance for bug reports at Q1, we see that in 25\% of the times that a code element is affected by a root-canal refactoring, it is necessary at least seven changes after the root-canal refactoring so that the code element becomes buggy, while that of floss requires at least eight changes for it to become buggy. Similar behavior is observed for the quartiles Q2 and Q3.
Surprisingly, when considering 100\% (Max) of the times that a code element affected by a root-canal refactoring becomes buggy, fewer changes (87 changes) are necessary after the application of a root-canal refactoring so that the code element becomes buggy compared to floss refactoring, which requires 364 changes. In this case, floss refactoring is less bug-prone compared to root-canal refactoring.

To analyze whether there is any statistically significant difference between the root canal and floss refactoring, we carried out the Mann-Whitney Wilcoxon test~\cite{conover1999}. The test returned a p-value less than 0.05, showing that there is a statistical difference between root-canal and floss refactoring when considering both insertion and report commits, with a confidence level of 95\%.

\begin{tcolorbox}[colframe=gray,colback=white,boxrule=2pt,arc=.3em,boxsep=-1mm]
    \textbf{RQ$_4$ Answer:} Our results suggest that generally, floss refactorings require more changes for the code to become buggy than root-canal refactoring. Hence, floss refactoring is less bug-prone than root-canal refactoring. 
\end{tcolorbox}

\subsection{RQ$_5$. How long does it take for a refactored code element to become buggy?}

Table~\ref{tab:distance_report_composite} presents the results of quartiles for the distance between the insertion and report commits related to a bug. The results are for both single and composite refactorings. We applied the Grubbs' test~\cite{grubbs1949}, considering a confidence level of 95\% (alpha = 0.05), and we removed outliers for both insertion and report commit data.

For bug reports, we noticed that in 25\% (Q1) of the time, a degraded code element takes at least seven changes to become buggy after single refactoring as compared to composite refactoring, which takes at least eight changes. Coincidentally, we found these seven and eight changes to have been performed in approximately three months between the single refactoring and the bug.
However, there is a notable variation when single and composite (see Tables~\ref{tab:root-canal-floss} and \ref{tab:distance_report_composite}) refactorings are compared at Q2, we see that in 50\% of the cases, single refactoring requires at least 19 changes for the refactored code to become buggy while composite refactoring requires about 29 changes. This is similar to the 75th percentile, where 42 changes for single refactoring to become buggy as compared to composite refactoring, which is at least 52 changes. This difference in a number of changes can also be seen for bug insertions when Q2, Q3, and Max columns are compared for bug insertions. Comparing this result to the results for RQ1, we can also say in general that single refactored code elements are more prone to become buggy than composite ones.

\begin{table}
\caption{Summary of Distance of Bug insertion and Report events for Composite Refactoring}
\label{tab:distance_report_composite}
\centering
\begin{tabular}{l c c c c c c}
\hline
\textbf{Bug} & \textbf{N} & \textbf{Min} & \textbf{Q1} & \textbf{Q2} & \textbf{Q3} & \textbf{Max} \\ \hline
Insertion    & 111         & 1.00         & 7.50        & 29.00       & 52.50       & 123.00       \\ 
Report       & 120         & 0.00         & 8.00        & 27.00       & 51.25       & 191.00       \\ \hline
\end{tabular}
\end{table}

\begin{tcolorbox}[colframe=gray,colback=white,boxrule=2pt,arc=.3em,boxsep=-1mm]
    \textbf{RQ$_5$ Answer:} Generally, it takes at least seven further changes for refactored code to become buggy. However, composite refactored code elements require more additional changes for them to become buggy than when single refactoring is applied to code elements. 
\end{tcolorbox}

Based on our results, and in response to the general question of our study ``\textit{What refactoring characteristics increase the likelihood of bugs in refactored code?}'', we conclude that certain factors contribute to higher bug proneness. These include the presence of code smells, specific refactoring types (i.e., Extract Method, Inline Method, Rename Method, Move Attribute, Move Method, and Extract Superclass), the application of single refactorings, mainly in root-canal refactoring. Together, these characteristics are indicators of code elements that may be more prone to bugs.


\section{Threats To Validity}
\label{sec:threats}
\subsection{Construct and Internal Validity}

The code smell types analyzed in this study might not be representative of all the existing varieties of code smells, so we selected the most common smell varieties that are directly related to refactoring types analyzed in this work~\cite{19, 21}. Also, our analysis of the code smells is sensitive to thresholds associated with detection rules. Hence, we used thresholds previously validated by other researchers~\cite{14, 33, 61, 62}.

The refactoring types we analyzed may not represent all types, and hence, we selected refactoring types that have been widely applied and investigated by previous studies~\cite{1, 6, 19}. We mitigate the effects of false positives introduced by the Refactoring Miner tool by manual validation to guarantee the reliability of our results.

Regarding the refactoring purposes, we could not reach the developers to ask about their intentions (i.e., root-canal or floss) in all detected refactorings. Therefore, we performed a manual validation to see whether the refactoring is root-canal or floss based on behavior presentation in the refactored code.

Lastly, our study was based on bugs reported or detected by tools, thus some bugs might be missed in our analysis. To mitigate the effect of incorrect tagging reports as ``bug'' or ``defect'', we again performed manual validation of a sample.

\subsection{Conclusion and External Validity}
Our selection of 12 projects may be a low number of instances for an empirical study of this nature, but we minimize possible threats to validity by analyzing software projects from different domains, with varying sizes, supported by active issue tracking systems and using of well-known statistical methods like \textit{quartiles}~\cite{31}.

\section{Conclusion And Future Work}
\label{sec:Conclusion}
In this paper, we investigated the bug-proneness of code refactored in the context of various refactoring characteristics.
Our study involved 12 Java open source projects with 27,450 refactorings (19,622 single refactorings and 7,828 composite refactorings), 6,051 bug reports, and 49,250 bugs detected by a static analysis tool.
Our results show that refactored code is not often susceptible to containing bugs, but by considering the distance property, we noticed that when a refactored code element is changed many times, that code element becomes buggy. Furthermore, our results prompt us to make the following recommendations:
\begin{itemize}
    \item  Single refactoring applied to code elements with smell tends not to suffice to reduce their bug-proneness. Hence, developers should apply complementary refactorings to fully remove the code degradation. Also, composite refactoring generally requires more changes to become buggy than single refactoring. Thus, developers should apply composite refactoring whenever possible.
    \item Developers should examine refactoring intent carefully in order to make an accurate decision on whether to apply floss or root-canal refactoring.
    
    \item The application of Extract Method and Inline Method refactorings must be done with caution since these refactoring types are susceptible to containing immediate bugs.

    \item Researchers may focus on developing a distance-aware bug prediction model for refactored code, leveraging historical refactoring patterns to enhance defect prediction and assist developers in mitigating risks.
\end{itemize}

In future work, we intend to assess the bug-proneness of regular changes, e.g., line additions. After that, we can compare the bug-proneness of refactored code against the bug-proneness of code elements that had regular changes. This analysis will allow us to see whether other change types make the source code bug-prone. Furthermore, it is interesting to assess the bug-proneness of refactored code in proprietary software systems. Our studies focused only on the analysis of popular open-source projects, which may have different structural degradation, different types of bugs, and refactorings as compared to proprietary software systems. Also, the exclusion of confounding factors like team dynamics and process complexity may impact our findings~\cite{6606589}. While our focus is on refactoring, this limitation could affect generalizability. Future studies may address this by incorporating these factors. Finally, we intend to recommend practices to motivate software developers to refactor and improve state-of-the-art refactoring tools.


\balance

\bibliographystyle{ACM-Reference-Format}
\bibliography{references}
\end{document}


\definecolor{cellh}{gray}{1.0} 

\newcommand{\la}[1]{{\textcolor{magenta}{~[~\textbf{LA}: \textit{#1} ]}}}
\newcommand{\wk}[1]{{\textcolor{red}{~[~\textbf{WK}: \textit{#1} ]}}}

\title{Assessing the Bug-Proneness of Refactored Code:\\A Longitudinal Multi-Project Study}

\author{\IEEEauthorblockN{Anonymous Author(s)}}

\maketitle
This document presents supplementary material for the paper “Assessing the Bug-Proneness of Refactored Code:\\A Longitudinal Multi-Project Study”.

\begin{table*}[h!]
\centering
\caption{Analyzed Refactoring Types}
\def\arraystretch{2.0}
\begin{tabular}{m{4cm} m{6cm} m{7cm}}
\toprule
\textbf{Refactoring Type} & \textbf{Problem}  & \textbf{Solution}  \\
\midrule
Extract Interface & Several clients use the same subset of a class’s
interface, or two classes have part of their interfaces in common & Extract the subset into an interface\\
\hline
Extract Method & A code fragment can be grouped together & Turn the fragment into a method whose name
explains the purpose of the method \\
\hline
Extract Superclass & There are two classes with similar features & Create a superclass and move the common
features to the superclass \\
\hline
Inline Method &  When a method body is more obvious than the
method & Replace calls to the method with the method’s
content and delete the method itself\\
\hline
Move Field  & A field is, or will be, used by another class more
than the class on which it is defined & Create a new field in the target class, and change
all its users\\
\hline
Move Class  & Your class belongs to a package that other classes
unrelated to it & Move the class to a related package or create a
new package if required for further use\\
\hline
Move Method  &  A method is, or will be, using or used by more
features of another class than the class in which
it is deffined & Create a new method with a similar body in the
class it uses most. Either turn the old method
into a simple delegation, or remove it altogether\\
\hline
Pull Up Field  &  Two subclasses have the same field & Move the field to the superclass \\
\hline
Pull Up Method  &  There are methods with identical results on
subclasses & Move them to the superclass \\
\hline
Push Down Field  &  A field is used only by some subclasses & Move the field to those subclasses\\
\hline
Push Down Method  &  The behavior on a superclass is relevant only for
some of its subclasses & Move it to those subclasses\\
\hline
Rename Class  &  The name of the class does not reveal its purpose & Change the name of the class and update all
callers\\
\hline
Rename Method  &  The name of a method does not reveal its purpose & Change the name of the method and update all
callers\\
\midrule
\end{tabular}
\label{tab:selected-projects}
\end{table*}

\begin{table*}[t!]
\centering
\caption{Analyzed Code Smell Types}
\def\arraystretch{2.0}
\begin{tabular}{m{4cm} m{13cm}}
\toprule
\textbf{Smell Type} & \textbf{Description} \\
        \hline
        Brain Class & Long and complex class that centralizes the intelligence of the system \\
        \hline
        Brain Method & Long and complex method that centralizes the intelligence of a class \\
        \hline
        Class Data Should Be Private & A class exposing its fields, violating the principle of data hiding \\
        \hline
        Complex Class & A class having at least one method with high cyclomatic complexity \\
        \hline
        Data Class & Classes that have only fields and accessor methods \\
        \hline
        Dispersed Coupling & A method that accesses many code elements, and the accessed elements are dispersed among many classes \\
        \hline
        Feature Envy & A method that is more interested in a class other than the one it actually is in \\
        \hline
        God Class & When a class centralizes the system functionality \\
        \hline
        Intensive Coupling & A method that has tight coupling with other methods, and these coupled methods are defined in the context of a few classes \\
        \hline
        Lazy Class & A class with very small dimensions, few methods, and low complexity \\
        \hline
        Long Method & A method that is unduly long in terms of lines of code \\
        \hline
        Long Parameter List & A method having a long list of parameters, some of which are avoidable \\
        \hline
        Message Chain & A long chain of method invocations is performed to implement a class functionality \\
        \hline
        Refused Bequest & A class redefining most of the inherited methods, signaling a wrong hierarchy \\
        \hline
        Shotgun Surgery & When a change demands many small changes across several different classes \\
        \hline
        Spaghetti Code & A class implementing complex methods that interact with each other, without parameters, using global variables \\
        \hline
        Speculative Generality & An abstract class with very few children classes using its methods \\
        \hline
    \end{tabular}
\end{table*}
